\documentclass[journal=pra, manuscriot=article]{revtex4}
\usepackage[utf8]{inputenc}
\usepackage{amssymb}
\usepackage{amsmath}
\usepackage{amsthm}
\usepackage{bm}
\usepackage{xcolor}
\usepackage{hyperref}

\renewcommand{\i}{\mathrm{i}}
\renewcommand{\d}{\mathrm{d}}
\newcommand{\xx}{\mathbf{x}}
\newcommand{\rr}{\mathbf{r}}

\newcommand{\bla}{\bm{\lambda}_\alpha}
\newcommand{\blb}{\bm{\lambda}_\beta}
\newcommand{\blg}{\bm{\lambda}_\gamma}
\newcommand{\qa}{\hat{q}_\alpha}
\newcommand{\qb}{\hat{q}_\beta}
\newcommand{\pa}{\hat{p}_\alpha}
\newcommand{\pb}{\hat{p}_\beta}
\newcommand{\oa}{\omega_\alpha}
\newcommand{\ob}{\omega_\beta}
\newcommand{\EX}[1]{\langle #1 \rangle}
\newcommand{\R}{\mathbb{R}}
\newcommand{\Hb}{\hat H_\mathrm{b}}

\newcommand{\Hc}{\hat H_\mathrm{c}}
\newcommand{\Hd}{\hat H_\mathrm{d}}
\newcommand{\Hext}{\hat H_\mathrm{ext}}

\begin{document}
\title{Virial relations for electrons coupled to quantum field modes}
\author{Iris Theophilou}
\affiliation
{Theory Department, Max Planck Institute for the Structure and Dynamics of Matter, Luruper Chaussee 149, 22761 Hamburg, Germany}
\email{iris.theophilou@mpsd.mpg.de}
\author{Markus Penz}
\affiliation
{Theory Department, Max Planck Institute for the Structure and Dynamics of Matter, Luruper Chaussee 149, 22761 Hamburg, Germany}
\author{Michael Ruggenthaler}
\affiliation
{Theory Department, Max Planck Institute for the Structure and Dynamics of Matter, Luruper Chaussee 149, 22761 Hamburg, Germany}

\author{Angel Rubio}

\affiliation{Theory Department, Max Planck Institute for the Structure and Dynamics of Matter, Luruper Chaussee 149, 22761 Hamburg, Germany}
\affiliation{Center for Computational Quantum Physics (CCQ), Flatiron Institute, 162 Fifth Avenue, New York NY 10010, USA}

\begin{abstract}

In this work we present a set of virial relations for many electron systems coupled to both classical and quantum fields, described by the Pauli--Fierz Hamiltonian in dipole approximation and using length gauge. Currently, there is growing interest in solutions of this Hamiltonian due to its relevance for describing molecular systems strongly coupled to photonic modes in cavities, and in the possible modification of chemical properties of such systems compared to the ones in free space. The relevance of such virial relations is demonstrated by showing a connection to mass renormalization and by providing an exact way to obtain total energies from potentials in the framework of Quantum Electrodynamical Density Functional Theory.

\end{abstract}

\maketitle

\section{Introduction}

Recent experimental progress allows to manipulate intrinsic properties of molecules by placing them inside photonic structures providing strong and ultrastrong light–matter coupling with a confined electromagnetic vacuum in the optical and infrared regimes\cite{Chemical_reactions_Ebbesen, Singe_quantum_emitter,Single_molecule_strong_coupling,suppression_photobleaching, Selective_manipulation_excited_states, energy_transfer, Lidzey1998}. Under such conditions, molecular polaritons are formed: hybrid energy states composed of entangled matter-photon degrees of freedom. As a result of the formation of polaritons, the matter system exhibits very different properties than in free space, opening numerous possibilities for manipulating chemical processes \cite{Ebbesen2016, Kockum2018, Herrera_Owrutsky_review,Review_Borjesson, Ruggenthaler2018}. These new hybrid polaritonic entities manifest themselves in numerous ways, including modifications of chemical reaction rates \cite{Ebbesen2016}, the possibility of long-range energy transfer between matter systems \cite{energy_transfer, Christian_charge_transfer} and the suppression of photobleaching \cite{suppression_photobleaching}, to give just a few examples.

In order to understand and eventually manipulate chemical reactions by strong coupling to confined light modes, the development of theoretical methods that are at the interface between quantum chemistry, solid state physics, and quantum optics have emerged \cite{Tokatly_QED, ruggenthaler2015groundstate, Ruggenthaler2014, Johannes_Response, flick2019excitedstate,OEP_QEDFT, OEP_ground_state, dressed_RDMFT, buchholz2020lightmatter, Suppression_of_isomerizaton,Cavity_Casmir_Polder, mordovina2019polaritonic,haugl2020coupled,
Variational_QED_Johannes,Abedi2018,Exact_factorization_Neepa,resonant_catalysis_Yuen-Zhou,reactivities_Pengfei,multiscale_md_polaritonic_chemistry,VENDRELL_MCTDH,Potential_crossings_diatomics,CMD_VUC_Nitzan_Subotnik,Herrera2016,Polariton_chemistry_Ribeiro,Ehrenfest_Norah,Cavity_Femtochemistry_Mukamel,Keeling_states_spectra_vibrational_polaritons,Fregoni2018,electron_transfer_Nitzan}.
However, since for large systems these methods always have to rely on some level of approximation, theoretical guidelines that help to assess whether an approximation is valid and thus ultimately predictive are of utmost importance. Our focus will be on such guidelines for methods that treat the matter part of the system like in electronic-structure theory \cite{Tokatly_QED, ruggenthaler2015groundstate, Ruggenthaler2014, Johannes_Response, flick2019excitedstate,OEP_QEDFT, OEP_ground_state, dressed_RDMFT,buchholz2020lightmatter, Suppression_of_isomerizaton,Cavity_Casmir_Polder, mordovina2019polaritonic,
haugl2020coupled,
Variational_QED_Johannes,  Abedi2018,Exact_factorization_Neepa, Potential_crossings_diatomics,  resonant_catalysis_Yuen-Zhou, reactivities_Pengfei,multiscale_md_polaritonic_chemistry, VENDRELL_MCTDH,CMD_VUC_Nitzan_Subotnik},
and not solely as a few level system. 

When it comes to electronic-structure problems, the currently most popular approach is Density Functional Theory (DFT) \cite{parr}. Several exact constraints that functionals of DFT should adhere to are known \cite{Perdew2003}. Such constraints have served as both, sanity checks for approximate functionals and contributions to the development of new ones \cite{Scan}. They even led to the improvement of functionals in combination with machine-learning techniques\cite{machine_learning_exact_conditions}. One of these exact constraints relates to the virial theorem, whose extension to the realm of electronic systems coupled to field modes is the main focus of this work. Recently, a virial-motivated model has been introduced, by which one can obtain accurate singlet-triplet splittings \cite{Becke_virial} and charge-transfer excitation energies \cite{Becke_virial_charge_transfer} from ground-state DFT calculations. Another recent application of the virial theorem was to provide a way to calculate the ``non-additive'' kinetic energy in terms of quantities that can all be obtained through embedding partition-DFT calculations \cite{Partition_DFT}. The virial theorem has been also used to calculate atomic energies in the context of Quantum Theory of Atoms In Molecules within the Kohn--Sham DFT formalism \cite{QTAIM}.
The role of the virial theorem is also important for other classes of quantum-chemistry methods that rely on approximate wave functions \cite{Jensen}, as it is used to assist geometry optimization \cite{wave_functions_satisfying_virial} in electronic-structure codes (e.g., \texttt{GAMESS (US)} \cite{Gamess_US}) and as an indicator of the basis-set quality \cite{virial_indicator_basis_set}.
Importantly, electronic-structure methods are assessed against exact Full Configuration Interaction or other accurate wave-function methods results, which are still numerically feasible for small molecules (e.g., \texttt{G2/97} and \texttt{G3/99} theoretical thermochemistry test sets). These results serve as a consistency check for the reliability of every newly developed method. However, when it comes to many-electron systems coupled to photon modes, obtaining such exact reference solutions becomes ever more demanding because of the increased dimensionality of the involved Hilbert spaces. Only very recently, exact calculations became available for
$\mathrm{HD^+}$ and $\mathrm{H_2^+}$ molecules as well as for the He-atom. \cite{sidler2020chemistry}. Consequently, theoretical guidelines become even more important.

Our physical starting point is the Pauli--Fierz Hamiltonian \citep{spohn2004dynamics}, which describes $N$ electrons coupled to quantized field modes.
We will restrict ourselves to the dipole approximation, which is valid for a big variety of situations where the spatial extension of the field modes is much larger than the matter system itself, and which is commonly used in cavity QED. Such a model is also useful to capture the effects of the quantized electromagnetic field in free space and thus connects to the old question of how quantum chemistry is affected by the electromagnetic vacuum. Obvious effects are the finite lifetime of excited states or the Lamb shift, a more subtle effect will be highlighted in Section~\ref{sec:mode-couple-virial-renorm} with a brief discussion on mass renormalization. The matter system will be treated within the usual  ``clamped-nuclei approximation'',  where the nuclei just contribute to a fixed external scalar potential $v(\rr)$, as it is the case in most electronic-structure approaches. On the other hand, we do not restrict the number of modes to just one effective mode, as it is commonly done in practice, but allow arbitrarily many modes in order to sample the photon continuum~\cite{Flick2017,flick2019excitedstate} and to be able to describe situations where many modes become relevant~\cite{PhysRevLett.117.153601}.Further, our results are not restricted to polaritonic ground states, but extend to any eigenstate of the polaritonic system.\par

Within this setting we present a set of virial relations between the energy expectation values of different parts of the Pauli--Fierz Hamiltonian in an eigenstate. Such relations can be derived by considering the time derivative of expectation values of different time-independent operators, which are zero if an eigenstate of the Hamiltonian is considered. By means of one such operator, consequently called a ``virial operator'', we obtain a virial relation that is the generalization of the quantum virial theorem for electronic structure \cite{Virial_Slater}. It will involve two additional terms that stem from the interaction of the electrons with the photons: the dipole coupling energy and the quadratic dipole self-energy.
We use the term ``virial'' for two more identities that we present here, signifying simply that these are exact relations that eigenstates of the system must satisfy and that involve one or more of the separate energy terms from the Hamiltonian.\par
Outline of the article: After introducing the physical setup of a system of interacting electrons coupled to photon modes in Section~\ref{sec:setup}, we give a general method of how to derive virial relations from equations of motion in stationary states in Section~\ref{sec:EOM-virial}. Three different virial operators are then used in the following three sections to arrive at different virial relations. First, a basic quantum virial theorem is derived in Section~\ref{sec:basic-virial}, in which the terms from the purely electronic part of the Hamiltonian are combined with the dipole coupling and dipole self-energy. The next virial theorem given in Section~\ref{sec:mode-virial} is then concerned with photon-field constituents and relates the parts of the photon-field energy to the dipole coupling.
Another virial relation that is only given as an estimate is derived in Section~\ref{sec:mode-couple-virial-renorm}. It allows to make an interesting connection to mass renormalization due to coupling to the field modes. Finally, in Section~\ref{sec:kohn-sham-energy}, the derived virial theorems are employed to map from potential functionals to energy functionals in the Kohn--Sham approach of Quantum Electrodynamical DFT (QEDFT). The paper is concluded in Section~\ref{sec:summary} with a short summary of the main results and a brief outlook. 

\section{Physical Setup}
\label{sec:setup}

A system of $N$ electrons with position-spin coordinates $\xx_i = (\rr_i,s_i)$, $\rr_i \in \R^d$, $s_i \in \{ \uparrow,\downarrow \}$, $i = 1,\ldots,N$ is coupled to the field modes with fundamental light-matter coupling $\bla$, which includes the effective coupling strength and polarization direction (for details see (7) in \citet{SchaeferRelevance}), and the mode frequency $\oa$ and mode coordinate $q_\alpha$, $\alpha = 1,2,\ldots$, where an infinite, countable number of modes is permitted. The nature of the field modes is not important here, they could either be due to a cavity environment or a free-field continuum. For the electrons we will always assume a position basis and thus write simply $\rr_i$ for the respective position operator. The gradient with respect to any coordinate is written $\nabla_i$ and the divergence is the inner product with $\nabla_i$, so the momentum operator is $-\i\nabla_i$ and the Laplacian $\nabla^2_i$. On the side of the bosonic field modes the coordinates are attached to the operators $\qa$ and their conjugates $\pa$, connected through the canonical commutation relation,
\begin{equation}\label{eq:comm-qp}
    [\qa,\pb] = \i \delta_{\alpha\beta}.
\end{equation}
Note that in the literature $\qa,\pa$ sometimes take exactly the opposite roles, but this just amounts to a relabelling. Physically the mode coordinate corresponds to the displacement field, while its conjugate represents the magnetic field.
Atomic units will be employed throughout the article, meaning that the elementary charge, electron mass, $\hbar$, and $4\pi\epsilon_0$ are all set to unity by definition. The system's Hamiltonian is given in length gauge\cite{Tokatly_QED,SchaeferRelevance}, 
\begin{equation}\label{eq:ham}
    \hat H = \hat T + \hat V + \hat W + \Hb + \Hc + \Hd + \Hext,
\end{equation}
and the different parts will be given in the following. We use atomic units throughout the manuscript. The kinetic part in length gauge contains only the canonical momentum $-\i\nabla_i$ and includes no reference to the mode coordinates,
\begin{equation}
    \hat T = -\frac{1}{2}\sum_i \nabla_i^2.
\end{equation}
The potential terms $\hat V, \hat W$ are the influence of an external scalar potential $v(\rr)$ and the usual Coulombic repulsion between the electrons,
\begin{equation}
    \hat V = \sum_i v(\rr_i), \quad \hat W = \frac{1}{2}\sum_{i\neq j} \frac{1}{|\rr_i-\rr_j|}.
\end{equation}
As the last term that depends exclusively on the electron coordinates, we have the quadratic dipole self-energy term,
\begin{equation}
    \Hd = \frac{1}{2} \sum_{i,j,\alpha} (\bla \cdot \rr_i) (\bla \cdot \rr_j) =  \frac{1}{2} \sum_{\alpha} \left(\bla \cdot \sum_i\rr_i \right)^2,
\end{equation}
with the coupling vectors $\bla \alpha \in
\mathbb{R}^d$. What follows is the dipole-coupling energy and the purely bosonic field energy in length gauge,
\begin{equation}
    \Hc = - \sum_{i,\alpha} \oa (\bla \cdot \rr_i) \pa, \quad
    \Hb = \frac{1}{2}\sum_\alpha \left(\oa^2\pa^2 + \qa^2\right).
\end{equation}
The last term is the coupling of the mode to an external force,
\begin{equation}
    \Hext = \sum_{\alpha} \frac{f_{\alpha}}{\omega_\alpha} \hat{p}_{\alpha},
    \label{eq:def_Hext}
\end{equation}
where we have $f_{\alpha}$ as the time-derivative of an external current \cite{Tokatly_QED}. Since some of the operators above are squares of self-adjoint operators, or sums of such squares, their expectation value is always positive. This positivity holds for $\EX{\hat T}, \EX{\hat W}, \EX{\Hd}, \EX{\Hb} \geq 0$.

Note that only $\hat T$ contains electron momentum operators, while $\hat V,  \hat W, \Hd$ contain only electron positions. On the other hand $\Hb$ and $\Hext$ contain only mode operators and $\Hc$ is the only part of the Hamiltonian that includes electron positions \emph{and} $\pa$. This will be important for commutators: If a commutator is between operators where the first includes only electronic coordinates and the second only mode coordinates, then it will automatically be zero.
Additionally to the canonical commutation relation \eqref{eq:comm-qp} we will use the following basic commutators,
\begin{align}
    &[\qa,\qb] = [\pa,\pb] = 0, \label{eq:comm-qq}\\
    &[\qa, \pb^2] = 2i \delta_{\alpha\beta} \pa, \label{eq:comm-qp2}\\
    &[\qa^2, \pb] = 2i \delta_{\alpha\beta} \qa. \label{eq:comm-q2p}
\end{align}

The Hamiltonian $\hat H$ allows a more compact form where we gather all terms that include mode coordinates, $\Hb, \Hc, \Hd$, into one field-energy part with a mode momentum $\pa$,
\begin{equation}\label{eq:H-transformed}
    \hat H = \hat T + \hat V + \hat W + \frac{1}{2}\sum_\alpha\left(\left(\oa\pa - \sum_{i} \bla \cdot \rr_i\right)^2 + \qa^2\right) + \Hext.
\end{equation}
Although not used further, this form is useful to see that $\pa$ corresponds to the displacement field of the mode, shifted by the electronic dipole, and that $\Hb + \Hc + \Hd$, made out only of squares of self-adjoint operators, is a positive operator and thus $\EX{\Hb}+\EX{\Hc}+\EX{\Hd} \geq 0$. For more details concerning the importance of including also the terms $\Hc$ and $\Hd$ see \cite{Rokaj_2018,SchaeferRelevance}. For a broad class of external potentials that include the usual molecular potentials, described in \citet[Section~X.2]{reed1975ii}, the whole $\hat H$ is then bounded below too and thus it is able to support a ground state. We will always assume in the following that $v(\rr)$ is from that class and is such that $\hat H$ has an eigenstate, $\hat H \Psi= E\Psi$.

Before we move to the main subject of this work, we want to point out that the following results do also hold when one considers classical instead of quantum fields. Classical field here means that the ground state in the photon-sector is fully determined by the $\EX{\pa}$. In this case we substitute the coupling term like
\begin{align}\label{eq:maxwell_substitution}
\Hc \longrightarrow - \sum_{i,\alpha} \oa \EX{\bla \cdot \rr_i} \pa  - \sum_{i,\alpha} \oa (\bla \cdot \rr_i) \EX{\pa} + \sum_{i,\alpha} \oa \EX{\bla \cdot \rr_i} \EX{\pa},  
\end{align}
where the last term is just a constant shift to not double-count the energy contribution due to the mean-field interactions. This allows for a factorization of the wave function in the matter and the photon sector and we have two coupled, non-linear equations. In the matter part we have to keep the dipole self-energy term $\Hd$ to still allow for eigenstates~\cite{Rokaj_2018}. The solution in the photon sector is a tensor product of harmonic-oscillator eigenstates for $\EX{\pa}=0$ and coherent states for $\EX{\pa} \neq 0$. Since only $\EX{\pa}$ enters, we can even replace the Schrödinger equation for the photon sector including the substitution \eqref{eq:maxwell_substitution} by the mode-resolved Maxwell's equation \eqref{eq:mode-force-balance}. This highlights that by the substitution \eqref{eq:maxwell_substitution} we actually consider the coupled Maxwell--Schr\"odinger problem~\cite{Ruggenthaler2018}. The virial relations derived in the subsequent work are therefore also valid if we consider classical fields.

\section{Equations of motion for virial relations}
\label{sec:EOM-virial}

The basic idea to arrive at virial relations is to use the simple result that the time derivative of the expectation value of any time-independent operator $\hat A$ is zero in an eigenstate of $\hat H$. To see this one could choose $\Psi$ such that the Schrödinger equation is
\begin{equation}\label{eq:SE}
    \i\dot\Psi = \hat H \Psi = E\Psi
\end{equation}
and evaluate the \emph{equation of motion}
\begin{equation}
    \frac{\d}{\d t} \EX{\hat A} = \frac{\d}{\d t} \langle \Psi, \hat A\Psi \rangle = \langle \dot\Psi, \hat A\Psi \rangle+\langle \Psi, \hat A\dot\Psi \rangle = \i E(\EX{\hat A} - \EX{\hat A}) = 0
\end{equation}
Here and in the following we use the short-hand notation $\EX{\hat A} = \langle \Psi, \hat A\Psi \rangle$ for the expectation value of an operator with respect to the state $\Psi$. It must not be noted separately which state $\Psi$ is meant, because we will always only consider one and the same exemplary eigenstate.
Since similarly we arrive from \eqref{eq:SE} at
\begin{equation}
    \frac{\d}{\d t} \EX{\hat A} = \i\langle [\hat H,\hat A] \rangle, \quad\quad \text{(Ehrenfest theorem)}
\end{equation}
it follows $\langle [\hat H,\hat A] \rangle = 0$ for eigenstates. This result is known as the ``hypervirial theorem'' \citep{hirschfelder1960classical}. Now $[\hat H,\hat A]$ could be all sorts of different operators for which we then know that the expectation value is zero, so if we arrive again at terms from the Hamiltonian \eqref{eq:ham}, we have found a virial relation. The difficulty lies in finding the appropriate operators $\hat A$.

\section{Electronic virial theorem with mode-coupling}
\label{sec:basic-virial}

The well-known quantum virial theorem for a system of electrons interacting via Coulomb repulsion and under the influence of an external scalar potential is \cite{Virial_Slater} 
\begin{equation}\label{eq:qu-virial}
    2\EX{\hat T} + \EX{\hat W} = N\langle \rr \cdot (\nabla v(\rr)) \rangle.
\end{equation}
Here $\rr=\rr_1, \nabla=\nabla_1$ and all the particles yield the same term $N$ times because of the assumed anti-symmetry of the wave function.
Only for special types of external potentials we also arrive back at the expectation value of $\hat V$. But take the monomial potential $v(\rr)=\rr^n$ and the virial theorem is $2\EX{\hat T}+\EX{\hat W}=n\EX{\hat V}$. The question that puts itself from the previous section is, which operator can be used such that $\langle [\hat H,\hat A] \rangle$ yields exactly the relation \eqref{eq:qu-virial}. The answer is the so called "virial operator" \cite{Bader_review} $\hat A = \sum_i \rr_i \cdot \nabla_i$  and we will demonstrate this with the full mode-coupled Hamiltonian \eqref{eq:ham} right away, where all parts except of the mode-only $\Hb$ contribute.
We will treat them one after another. Note that in a double sum over all particles, only the $i=j$ part will remain, because $\rr_i \cdot \nabla_i$ only affects the $i$-th particle coordinate. Thus,
\begin{equation}
    \sum_i [\hat T,\rr_i \cdot \nabla_i] = -\frac{1}{2} \sum_{i,j} [\nabla_j^2,\rr_i] \cdot \nabla_i = -\sum_j \nabla_j^2 = 2\hat T,
\end{equation}
where $[\nabla_j^2,\rr_i]$ means that the Laplacian is applied to each coordinate component operator separately which then makes up a new vector. Note that all terms are treated as operators, so $\nabla f(\rr) = (\nabla f(\rr)) + f(\rr)\nabla$ by the chain rule.
\begin{align}
    &\begin{aligned}
        \sum_i [\hat W,\rr_i \cdot \nabla_i] &= \frac{1}{2} \sum_{i,j\neq k} \rr_i \cdot [|\rr_j-\rr_k|^{-1},\nabla_i] \\
        &= \frac{1}{2} \sum_{j\neq k} \left( \rr_j \cdot (-\nabla_j |\rr_j-\rr_k|^{-1}) + \rr_k \cdot (-\nabla_k |\rr_j-\rr_k|^{-1}) \right) \\
        &= \frac{1}{2} \sum_{j\neq k} \left( \rr_j \cdot \frac{\rr_j-\rr_k}{|\rr_j-\rr_k|^{3}} - \rr_k \cdot \frac{\rr_j-\rr_k}{|\rr_j-\rr_k|^{3}} \right) \\
        &= \frac{1}{2} \sum_{j\neq k} \frac{1}{|\rr_j-\rr_k|} = \hat W
    \end{aligned}\\
    &\sum_i [\hat V,\rr_i \cdot \nabla_i] = \sum_{i,j} \rr_i \cdot [v(\rr_j), \nabla_i] = -\sum_i \rr_i \cdot (\nabla_i v(\rr_i)) \\
    &\sum_i [\Hc,\rr_i \cdot \nabla_i] = -\sum_{i,j,\alpha} \oa \rr_i \cdot [(\bla \cdot \rr_j), \nabla_i] \pa = \sum_{j,\alpha} \oa (\rr_j \cdot \bla) \pa = -\Hc \\
    &\sum_i [\Hd,\rr_i \cdot \nabla_i] = \frac{1}{2} \sum_{i,j,k,\alpha} \rr_i \cdot [(\bla \cdot \rr_j)(\bla \cdot \rr_k), \nabla_i] = -2\Hd
\end{align}

Taking all these results together the new electronic virial theorem with mode-coupling is found,
\begin{equation}\label{eq:qu-virial-modes}
    2\EX{\hat T} + \EX{\hat W} - \EX{\Hc} - 2\EX{\Hd} = N\langle \rr \cdot (\nabla v(\rr)) \rangle.
\end{equation}
We see that all terms from the original Hamiltonian are involved, \emph{except} the field-energy of the cavity modes $\Hb$ and the energy from the external force on the modes $\Hext$. Thus this virial relation connects all the constituents of the system which could be expressed of purely electronic degrees of freedom with the dipole-coupling energy which is the only term of mixed electron-boson nature.

Some comments are in order here. First, by setting the fundamental light-matter coupling to $\bla=0$, we recover the virial for electronic structure \eqref{eq:qu-virial} as requested.
An alternative route to derive the basic virial theorem is from the force-balance equation for stationary states \cite{Tokatly_QED}. If the force totals to $\bm{f}(\rr)=0$ at every point $\rr$, as will be the case for eigenstates, then taking the space integral $\int \bm{f}(\rr)\cdot\rr \,\d\rr$ results in \eqref{eq:qu-virial-modes}. Let us also point out that the implications of the force balance equation for functional construction in the context of time-dependent QEDFT \cite{Tokatly_QED, Ruggenthaler2014, OEP_QEDFT} has been discussed in \citet{Tokatly_2018}.  

\section{Field-mode virial theorem}
\label{sec:mode-virial}

Next we derive an analogous equation of motion on the side of the modes, which means that we combine the mode operators into an mode virial operator $\sum_\alpha \qa\pa$. In the equation of motion $\sum_\alpha \langle[\hat H,\qa\pa]\rangle$ only three terms contribute and we use the commutators \eqref{eq:comm-qp}, \eqref{eq:comm-qq}, \eqref{eq:comm-qp2}, and \eqref{eq:comm-q2p} for evaluation.
\begin{align}
    &\sum_\alpha [\Hb,\qa\pa] = \frac{1}{2} \sum_{\alpha,\beta} \left( \oa^2[\pb^2,\qa]\pa + \qa [\qb^2,\pa] \right) = \i \sum_\alpha ( -\oa^2\pa^2 + \qa^2) \label{eq:virial-ho}\\
    &\sum_\alpha [\Hc,\qa\pa] = -\sum_{i,\alpha,\beta} \ob (\blb \cdot \rr_i) [\pb,\qa] \pa = -\i \Hc\\
    &\sum_\alpha [\Hext,\qa\pa] = \sum_{\alpha,\beta} \frac{f_\beta}{\ob} [\pb,\qa] \pa = -\i \Hext
\end{align}
The relation \eqref{eq:virial-ho} yields, as one would expect, just the virial theorem for the harmonic oscillator. Together with the other relations, that just give back the coupling and external-force energies, we arrive at the field-mode virial theorem,
\begin{equation}\label{eq:mode-virial}
    \sum_\alpha \left( \oa^2 \langle \pa^2 - \qa^2 \rangle \right) - \EX{\Hc}  = \EX{\Hext}.
\end{equation}
We see that the field-mode virial theorem has a similar structure than the electronic virial theorem from \eqref{eq:qu-virial-modes}, with an external influence on the right and a connection between purely bosonic parts of the system with the dipole-coupling energy on the left. The field-mode virial theorem \eqref{eq:mode-virial} is also a consequence of the equation of motion that involves the second time-derivative of $\Hb$.

Another useful relation arises if we choose the equation of motion for the much simpler operator $\pa$. Here only $\Hb$ contributes and, using \eqref{eq:comm-q2p}, $[\hat H,\pa] = \i\qa$. Computing the second time derivative we have to look at the double commutator $[\hat H, [\hat H,\pa]]$, so
\begin{align}
    &[\Hb,\i\qa] = \frac{1}{2} \sum_{\beta} \oa^2[\pb^2,\qa] = \oa^2\pa\\
    &[\Hc,\i\qa] = -\i\sum_{i,\beta} \ob (\blb \cdot \rr_i) [\pb,\qa] = -\sum_{j} \oa (\bla \cdot \rr_j)\\
    &[\Hext,\i\qa] = \i\sum_{\beta} \frac{f_\beta}{\ob} [\pb,\qa] = \frac{f_\alpha}{\oa}
\end{align}
Summing those terms and taking the expactation value that must be zero, we have
\begin{equation}\label{eq:mode-force-balance}
    \oa^2 \EX{\pa} - \sum_{j} \oa \EX{\bla \cdot \rr_j} = -\frac{f_\alpha}{\oa}.
\end{equation}
This relation corresponds to the mode resolved Maxwell equation for the displacement field, see for example \citet{Johannes_Response}. In contrast to the connection between the electronic force-balance equation and the electronic virial theorem in the previous section, it does not give rise to a virial theorem by itself, but it will be employed in the derivation of the next virial relation.

\section{Mode-coupled virial estimate and connection to mass renormalization}
\label{sec:mode-couple-virial-renorm}

In the search for another virial relation that will relate electronic and field parts of the Hamiltonian, we consider the equation of motion of an operator that includes both, electron coordinates and mode coordinates. A promising candidate, as we will see, is the mixed virial operator $\sum_{i} (\bla \cdot \rr_i) \qa$.
We derive the non-zero commutators with the individual parts of $\hat H$ using the basic commutation relations \eqref{eq:comm-qp}, \eqref{eq:comm-qq}, and \eqref{eq:comm-qp2}.
\begin{align}
    &\sum_{i} [\hat T,\bla \cdot \rr_i] \qa = -\frac{1}{2} \sum_{i,j} [\nabla_j^2,\bla \cdot \rr_i] \qa = -\sum_{i} (\bla \cdot \nabla_i) \qa \label{eq:comm-T-bla-r-p}\\
    &\sum_{i} (\bla \cdot \rr_i) [\Hb,\qa] = \frac{1}{2} \sum_{i,\beta} \ob^2 (\bla \cdot \rr_i) [\pb^2,\qa] = -\i \oa^2 \sum_{i} (\bla \cdot \rr_i) \pa \\
    &\sum_{i} (\bla \cdot \rr_i) [\Hc,\qa] = -\sum_{i,j,\beta} \ob (\bla \cdot \rr_i) (\blb \cdot \rr_j) [\pb,\qa] = \i\oa \sum_{i,j}  (\bla \cdot \rr_i) (\bla \cdot \rr_j) \\
    &\sum_{i} (\bla \cdot \rr_i) [\Hext,\qa] = \sum_{i,j,\beta} \frac{f_\beta}{\ob} (\bla \cdot \rr_i) [\pb,\qa] = -\i\frac{f_\alpha}{\oa}\sum_{j} (\bla \cdot \rr_j)
\end{align}
This already looks promising, since the constituents of $\Hc$ and $\Hd$ reappear. To make this connection explicit we add the terms above, remember that the expectation value of their sum in an eigenstate is always zero, divide by $\i\oa$ and sum over $\alpha$. The resulting expectation value must still be zero,
\begin{equation}\label{eq:Hc-Hd-virial}
    \sum_{j,\alpha} \i\oa^{-1} \langle (\bla \cdot \nabla_j) \qa \rangle + \EX{\Hc} + 2\EX{\Hd} + \sum_{j,\alpha}\frac{f_\alpha}{\oa^2} \langle \bla \cdot \rr_j \rangle = 0.
\end{equation}
The rightmost term can be re-expressed with \eqref{eq:mode-force-balance} and the defining equation \eqref{eq:def_Hext} as,
\begin{equation}\label{eq:ext-force-sum}
    \sum_{j,\alpha}\frac{f_\alpha}{\oa^2} \EX{\bla \cdot \rr_j} = \EX{\Hext}+ \sum_\alpha \frac{f_\alpha^2}{\oa^4} .
\end{equation}

The term $\EX{\Hc} + 2\EX{\Hd}$, relating to dipole coupling and dipole self-energy, also already appears in the basic virial theorem \eqref{eq:qu-virial-modes}, just the remaining first term of \eqref{eq:Hc-Hd-virial} appears a bit unwieldy if we want to re-express it by parts of the length-gauge Hamiltonian. Indeed it corresponds exactly to the coupling term in velocity gauge, cf.~\citet[(10)]{Rokaj_2018} or, in other words, to the Fourier transform of $\Hc$ in both, electronic and mode coordinates. We will handle it further by deriving an estimate. Because $-\i\oa^{-1}\bla \cdot \nabla_j$ and $\qa$ are both self-adjoint, and so is their sum, their sum squared is a positive operator, and since they commute we get
\begin{equation}
    \langle (-\i\oa^{-1}\bla \cdot \nabla_j + \qa)^2 \rangle = \oa^{-2}\langle (-\i\bla \cdot \nabla_j)^2 \rangle + \langle \qa^2 \rangle - 2\i\oa^{-1} \langle (\bla \cdot \nabla_j) \qa \rangle \geq 0.
\end{equation}
This is the point where an estimate enters the virial relation that we will derive in this section, needed to handle the first term in \eqref{eq:Hc-Hd-virial} in terms of expressions that come directly from parts of the Hamiltonian $\hat H$,
\begin{equation}\label{eq:estimate-bla-nabla}
    \sum_{j,\alpha}\frac{1}{2} \oa^{-2}\langle (-\i\bla \cdot \nabla_j)^2 \rangle + \frac{N}{2}\sum_{\alpha}\langle \qa^2 \rangle \geq \sum_{j,\alpha}\i\oa^{-1} \langle (\bla \cdot \nabla_j) \qa \rangle.
\end{equation}
To proceed and derive an expression for the first sum above, we have to choose a certain distribution of modes. This could be one that privileges a specific direction by the choice of $\bla$, like in cavity QED, but let us
assume that the modes are those of free space, evenly distributed in $k$-space in $d=3$ dimensions. To every mode we assign a vector $\bla \in R^3$, $|\bla| = \lambda = \sqrt{4\pi/L^3}$, see \citet{SchaeferRelevance}, and let always two modes occupy a $k$-space volume of $(\omega_0/c)^3$, where $\omega_0=2\pi c/L$ is the lowest frequency in a cube with length $L$. We take two modes per $k$-space box because of the two different polarization directions of the photons.
If we now choose a radius $k=\oa/c$, then the number of modes within a thin spherical shell with thickness $\d k$ is $2\cdot(\omega_0/c)^{-3} 4\pi k^2 \d k$. For any mode in the shell with $\bla$ there are two other modes with $\blb,\blg$ such that they are all orthogonal to each other and thus give
\begin{equation}\label{eq:nabla-3-sum}
    \oa^{-2}\left( \langle (-\i\bla \cdot \nabla_j)^2 \rangle + \langle (-\i\blb \cdot \nabla_j)^2 \rangle + \langle (-\i\blg \cdot \nabla_j)^2 \rangle \right) = -\frac{\lambda^2}{\oa^2} \langle \nabla_j^2 \rangle.
\end{equation}
The specific orientations of $\bla,\blb,\blg$ do not matter at this point because the Laplacian $\nabla_j^2$ is spherical symmetric. The above also tells us that for any \emph{three} modes within the considered shell volume we get one contribution like above. Thus summing over all modes up to a cut-off $k \leq \Lambda$ we have for the complete sum $\mu = \sum_\alpha \oa^{-2}\lambda^2$ when passing to an integral \begin{equation}
    \mu =  \int_0^\Lambda \frac{\lambda^2}{(ck)^2} \frac{8 \pi k^2}{3(\omega_0/c)^3} \d k = \frac{8\pi c \lambda^2\Lambda}{3\omega_0^3} = \frac{4\Lambda}{3\pi c^2}.
\end{equation}
This value corresponds to the mass renormalization discussed in \citet[(3.41)]{Hainzl_Mass_Renorm}. Some comments on the cutoff $\Lambda$ introduced here are in order. Although somehow arbitrary, it is a standard ingredient of non-relativistic QED theory, as the bare mass of the electron is given with respect to such a cutoff. Among other choices, one could choose for example to allow for field modes with energy smaller than m$c^2$, since for larger energies the non-relativistic theory is anyway invalid. Such an ultraviolet cutoff is not expected to affect the low-energy processes considered in the setting discussed here.  
Rewriting the virial relation \eqref{eq:Hc-Hd-virial} with the estimate just derived we have,
\begin{equation}\label{eq:iso-mode-coupled-virial-relation}
    \mu \EX{\hat T} + \frac{N}{2}\sum_\alpha \EX{\qa^2} + \EX{\Hc} + 2 \EX{\Hd} + \EX{\Hext} + \sum_\alpha \frac{f_\alpha^2}{\oa^4} \geq 0.
\end{equation}
Usually the mass renormalization is derived by requiring the same dispersion relation for the bare electron and an electron coupled to the free-space modes. Here we have a field-mode virial relation where the kinetic energy of the electrons enters with a small prefactor that relates to change of electron mass due to the coupling to the free-space modes. A different distribution of modes would just lead to a different mass renormalization prefactor $\mu$, while the relation \eqref{eq:iso-mode-coupled-virial-relation} would stay exactly the same.

\section{QEDFT Kohn--Sham Potentials to energies via the virial theorem}
\label{sec:kohn-sham-energy}

As an application which highlights the usefulness of the virial relations we show in this section how they can be used to recover energies from potentials in the ground state Kohn--Sham approach to QEDFT \cite{ruggenthaler2015groundstate,OEP_ground_state}. This has special significance in cases where the energy functional is previously unknown because only the Kohn--Sham potential is provided, like in method based on the force-balance equation\cite{force_balance}.
In the usual Kohn--Sham approach to QEDFT, the coupled \emph{and} interacting system of interest is goverened by the full Hamiltonian \eqref{eq:ham}. Since calculating the ground state $\Psi$ in such a case is a formidable task for realistic systems including many particles and many field modes, as mentioned already one solves instead an auxiliary system with $\hat H^s$ that includes no electron interactions and has uncoupled modes, $\bla^s=0$. 
\begin{equation}
 \hat H^s=\hat T+ \hat V^s + \Hb + \Hext^s.   
\end{equation}
The auxiliary system $\hat H^s$ has forces acting on the modes $f_\alpha^s$ and includes an external potential $v^s(\rr)$, exactly in such a way that the ground state $\Phi$ of this system has the same one-particle density, $\rho(\bm{r})=\sum_i\EX{ \delta (\bm{r}-\bm{r}_i)}_{\Psi}=\sum_i\EX{ \delta (\bm{r}-\bm{r}_i)}_{\Phi}$,
as well as the same mode coordinate $p_\alpha = \EX{\pa}_{\Psi}=\EX{\pa}_{\Phi}$. We can thus write down the virial theorems \eqref{eq:qu-virial-modes} and \eqref{eq:mode-virial} for both systems.
\begin{equation}\def\arraystretch{2.0}
\begin{array}{ll}
    2\EX{\hat T}_\Psi + \EX{\hat W}_\Psi - \EX{\Hc}_\Psi - 2\EX{\Hd}_\Psi = \int \rr \cdot (\nabla v(\rr)) \rho(\rr) \d \rr \quad
    &\sum_\alpha \left\langle \oa^2\pa^2 - \qa^2 \right\rangle_\Psi - \EX{\Hc}_\Psi  = \sum_{\alpha} \frac{f_{\alpha}}{\omega_\alpha} p_{\alpha}
    \\
    2\EX{\hat T}_\Phi + \EX{\hat W}_\Phi = \int \rr \cdot (\nabla v^s(\rr)) \rho(\rr) \d \rr
    &\sum_\alpha \left\langle \oa^2\pa^2 - \qa^2 \right\rangle_\Phi = \sum_{\alpha} \frac{f^s_{\alpha}}{\omega_\alpha} p_{\alpha}
\end{array}
\end{equation}
Subtraction of the corresponding equations leads to
\begin{align}
&2\left(\EX{\hat T}_\Psi-\EX{\hat T}_\Phi\right) + \EX{\hat W}_\Psi - \EX{\Hc}_\Psi - 2\EX{\Hd}_\Psi = -\int \rr \cdot \left(\nabla (v^s(\rr)-v(\rr))\right) \rho(\rr) \d \rr,\\
&\sum_\alpha \left( \left\langle \oa^2\pa^2 - \qa^2 \right\rangle_\Psi - \left\langle \oa^2\pa^2 - \qa^2 \right\rangle_\Phi \right) - \EX{\Hc}_\Psi = - \sum_{\alpha} \frac{f^s_{\alpha}-f_\alpha}{\omega_\alpha} p_{\alpha}.
\label{eq:QEDFT_2}
\end{align}
Let us first stress that when looking only at the electronic part by setting $\bla=0$ and by identifying the Hartree interaction contribution, which depends just on the density, we can recover the known virial relation for Kohn--Sham DFT~\cite{Virial_DFT, Burke2006}. Clearly, there will be other relations apart from the virial, like scaling relations~\cite{Virial_DFT}, that the exact QEDFT functional should satisfy, which would be interesting to investigate in the future. More importantly, the combined virial relations for the full and the auxiliary system show a way how to recover the energies as functionals of $\rho(\rr)$ and $p_\alpha$ on the left from the Kohn--Sham potentials $v^s(\rr)-v(\rr)$ and $f^s_{\alpha}-f_\alpha$ on the right. The coupling energy has a special role in this, since it is the only part that appears in both equations and can thus be fully eliminated if functional expressions for the other terms are given. Interestingly, \eqref{eq:QEDFT_2} shows that one could obtain a functional for the coupling energy $\EX{\Hc}_\Psi$ by approximating the term $\left\langle \oa^2\pa^2 - \qa^2 \right\rangle_\Psi$ by an expression that depends on the density (such an expression we could obtain from a reference calculation of, e.g., a uniform electron gas coupled to a cavity-field mode) in combination with \eqref{eq:mode-force-balance}, which would give us an expression for $f^s_{\alpha}$ for a given $p_{\alpha}$.

\section{Summary and Outlook}
\label{sec:summary}

We have derived three virial relations for the Pauli--Fierz Hamiltonian in dipole approximation. The first one in \eqref{eq:qu-virial-modes} we call electronic virial with mode-coupling and it involves the electronic parts of the Hamiltonian as well as the dipole-coupling energy and the dipole self-energy,
\begin{equation*}
    2\EX{\hat T} + \EX{\hat W} - \EX{\Hc} - 2\EX{\Hd} = N\langle \rr \cdot (\nabla v(\rr)) \rangle.
\end{equation*}
For zero coupling $\bla=0$ we recover the electronic virial theorem as the dipole-coupling energy $\EX{\Hc}$ and the dipole-self energy $\EX{\Hd}$ are zero. By means of the above relation the dipole-coupling energy $\EX{\Hc}$ can be re-expressed in terms of only electronic degrees of freedom in every eigenstate. Next we have presented a field-mode virial theorem that involves the bosonic parts of our system as well as the dipole-coupling energy $\EX{\Hc}$,
\begin{equation*}
    \sum_\alpha \left( \oa^2 \langle \pa^2 - \qa^2 \rangle \right) - \EX{\Hc}  = \EX{\Hext}.
\end{equation*}
We see that $\EX{\Hc}$ can be re-expressed again, this time in terms of only bosonic degrees of freedom in every eigenstate. As a corollary we show that combining these two virial theorems we get a new one that connects the purely electronic (on the left side) with the purely field-mode parts (on the right side) of the Hamiltonian and includes no mixed terms,
\begin{equation}
    2\EX{\hat T} + \EX{\hat W} - 2\EX{\Hd} - N\langle \rr \cdot (\nabla v(\rr)) \rangle = \sum_\alpha \left( \oa^2 \langle \pa^2 - \qa^2 \rangle \right)  - \EX{\Hext}.
\end{equation}
The equation above provides an energy balance relation between the fermionic and the photonic subsystem.
As a first interesting example, we have demonstrated in Section~\ref{sec:mode-couple-virial-renorm}, how a virial estimate derived in a similar manner as the other two virial theorems leads to a connection to mass renormalization due to the coupling to free-space modes.
 This example also showed how to explicitly use the free-space distribution of modes to calculate the influence of the vacuum on the virial relation. This highlights the opportunity to investigate the influence of the usually neglected free-space quantum vacuum on chemical properties. As another example of the relevance of our results, we have demonstrated how by means of our virial theorems with mode-coupling one could for given auxilary potentials in the context of QEDFT reproduce the coupling energy term in the Hamiltonian. This would be especially useful for functionals that are not based on functional derivatives of energy expressions. Besides this, the obtained virial relations offer guidance for the development of approximate exchange-correlation potentials in QEDFT that have to model the electron-electron as well as the electron-photon interaction terms. While for the electron-electron interaction many approximations are available, this is not the case for the electron-photon interaction or for the photon-photon interaction mediated by matter. The obtained relations allow to connect and test approximations in the different sectors. 
Let us also point out that the method to obtain virial relations from Section~\ref{sec:EOM-virial} can lead to precisely the same result for different ``virial operators''.
This said, it is still possible that other choices of operators lead to additional useful ``virial relations'' that need to be satisfied if the system is in an eigenstate. Such results lead to a hierarchy of constrains that the system needs to fulfill and that can serve as a useful tool in analyzing the stationary properties of complex quantum systems coupled to the quantized electromagnetic field.

\section{Acknowledgement}
I.T. would like to thank Florian Buchholz and Vasil Rokaj for  stimulating and useful discussions. This work was supported by the European Research Council (ERC-2015-AdG694097), the Cluster of Excellence ``Advanced  Imaging of Matter'' (AIM) and SFB925. The Flatiron Institute is a division of the Simons Foundation.

\section*{References}
\bibliographystyle{apsrev4}
\bibliography{draft_new}
\end{document}